\documentclass[10pt,journal,comsoc,letterpaper,oneside,twocolumn]{IEEEtran}
\usepackage[T1]{fontenc}% optional
\usepackage{amsfonts}
\usepackage{amsmath}
\usepackage[cmintegrals]{newtxmath}
\usepackage{bm}% optional
\usepackage{setspace}
\usepackage{algorithm}
\usepackage{algpseudocode}
\usepackage{listings}
\usepackage{graphicx}
\usepackage[caption=false,font=footnotesize]{subfig}
\usepackage{cite}
\usepackage{verbatim}
\usepackage{makecell}
\usepackage{booktabs}
\usepackage{tikz,xcolor,hyperref}% Make Orcid icon
\newtheorem{defn}{Definition}
\newtheorem{lem}{Lemma}
\newtheorem{rem}{Remark}

\newtheorem{thm}{Theorem}
\newtheorem{assum}{Assumption}
\renewcommand\thetable{\Roman{table}}
\definecolor{lime}{HTML}{A6CE39}
\DeclareRobustCommand{\orcidicon}{%
    \begin{tikzpicture}
    \draw[lime, fill=lime] (0,0) 
    circle [radius=0.16] 
    node[white] {{\fontfamily{qag}\selectfont \tiny ID}};    \draw[white, fill=white] (-0.0625,0.095) 
    circle [radius=0.007];    \end{tikzpicture}
    \hspace{-2mm}}
\foreach \x in {A, ..., Z}{%
    \expandafter\xdef\csname orcid\x\endcsname{\noexpand\href{https://orcid.org/\csname orcidauthor\x\endcsname}{\noexpand\orcidicon}}
    }

\hypersetup{hidelinks}
\ifCLASSINFOpdf
\else
\fi
\begin{document}
\title{Online Adaptive Optimal Control Algorithm Based on Synchronous 
Integral Reinforcement Learning With Explorations}

\author{Lei~Guo\orcidA{},~\IEEEmembership{Member,~IEEE},
        and Han~Zhao\orcidB{},~\IEEEmembership{Student~Member,~IEEE}
\thanks{This work was supported by National Natural Science Foundation of China under Grant 61105103. \emph{(Corresponding author: Lei Guo.)}} % <-this % stops a space
\thanks{The authors are with the School of Artificial Intelligence, Beijing University of Posts and
Telecommunications, P.O. Box 108, No.10 Xi Tucheng Road, Haidian, Beijing 100876, China 
(e-mail: guolei@bupt.edu.cn; han\_zhao@bupt.edu.cn).}}% <-this % stops a space% <-this % stops a space


\markboth{PREPRINT SUBMITTED TO ARXIV.ORG} %
{Guo \MakeLowercase{\textit{et al.}}: Online adaptive optimal control algorithm based on synchronous 
integral reinforcement learning with explorations}



\maketitle

\begin{abstract}
In this paper, we present a novel algorithm named synchronous integral 
Q-learning, which is based on synchronous policy iteration, to solve 
the continuous-time infinite horizon optimal control problems 
of input-affine system dynamics. The integral reinforcement is measured 
as an excitation signal in this method to estimate the solution to the 
Hamilton--Jacobi--Bellman equation. Moreover, the proposed method 
is completely model-free, i.e. no\emph{ a priori} knowledge of the 
system is required. Using policy iteration, the actor and 
critic neural networks can simultaneously approximate the optimal value function 
and policy. The persistence of excitation condition is required 
to guarantee the convergence of the two networks. Unlike in 
traditional policy iteration algorithms, the restriction of the initial admissible 
policy is relaxed in this method. The effectiveness of the proposed 
algorithm is verified through numerical simulations.
\end{abstract}

\begin{IEEEkeywords}
Synchronous integral reinforcement learning, Policy iteration, Persistence of excitation, Adaptive control.
\end{IEEEkeywords}

% \fi
\IEEEpeerreviewmaketitle

\section{Introduction}
\IEEEPARstart{O}{ptimal} control \cite{Lewis1995} and adaptive control \cite{Ioannou2006} 
are two important concepts in modern control theory. The main goal 
of the optimal/adaptive controller is to reach the control objective 
with the minimal performance index/the unknown system structures 
or parameters. The method that combines the advantages of both 
methods is called reinforcement learning (RL, \cite{Sutton1998}) in the
computational intelligence field or adaptive dynamic programming 
(ADP, \cite{Werbos1974}) in control theory (also known as approximate 
dynamic programming \cite{Werbos1977}, neuro-dynamic programming 
\cite{Bertsekas1996} and adaptive critic design \cite{Prokhorov1997}), 
and it has been widely studied (See \cite{Liu2021} for the latest survey 
on ADP).

The key problem of optimal control/ADPRL methods is how to solve  
the Hamilton--Jacobi--Bellman equation (HJBE) or Bellman equation, which 
is the discrete-time (DT) version of the HJBE and often used in the RL 
literature. The optimal policy and the corresponding representation 
of its quality, i.e. the value function (VF), can be solved from the HJBE. \
Owing to the phenomenon known as “curses 
of dimensionality” \cite{Bellman1957}, the exact solution of the HJBE 
is usually difficult to find. The approximation method is often used, e.g. 
iterative methods using neural networks (NNs) \cite{AbuKhalaf2005}. 
The well-known actor-critic structure is generally used in ADPRL 
methods to simultaneously approximate the optimal policy and its VF.

Meanwhile, model-based methods may be difficult to implement 
in real-world control problems owing to the difficulties in mechanism 
modelling and the uncertainties of the dynamics system, which are 
called “curses of modelling” \cite{Bertsekas1996}. In studies on DT 
Markov decision process, model-free methods 
in ADPRL and deep RL based on deep NNs have achieved considerable 
success \cite{Watkins1989,Lillicrap2016,Schulman2017}. In the continuous-time 
(CT) domain, however, the effective methods in DT systems, e.g. 
action-dependent heuristic dynamic programming \cite{Werbos1989} 
or Q-learning \cite{Watkins1989}, are difficult to implement 
because\emph{ a priori} knowledge and partial difference 
forms are required in the CT HJBE. \cite{Baird1994} proposed an advantage 
updating algorithm to approximately compute the derivative of the 
VF. A model-free estimation method of the VF was also proposed in 
\cite{Murray2002}; however, an approximation or measurement of the differential 
term in these two methods is needed.

To solve the aforementioned problem, \cite{Vrabie2009} proposed the 
concept of integral RL (IRL) and an algorithm to solve the CT 
optimal control problem of linear systems. The temporal 
difference (TD, \cite{Sutton1988}) estimation was introduced into the 
IRL algorithms by solving the integral form of the HJBE. 
The requirement that the system dynamics must be fully known is 
relaxed in \cite{Vrabie2009}. Under the persistence of excitation 
(PE, \cite{Ioannou2006}) condition, the VF of the current policy can be 
estimated in a model-free manner, and the drift dynamics of the system are 
not used in policy updates. However, satisfaction with the PE requirement 
cannot be guaranteed during the estimation of the parameters. \cite{Jiang2012} 
added the exploration signal to the input to excite the system and  
removed the restriction of the\emph{ a priori} knowledge of the input gain 
matrix. For nonlinear problems, \cite{Lee2012} used an exploration 
method that was extended and improved in \cite{Lee2015}.

These aforementioned IRL algorithms are based on policy iteration (PI), which 
is an iterative method of dynamic programming (DP). The latest summary of 
the PI algorithms in ADP 
field can be found in \cite{Lee2021}. To guarantee the convergence of 
the weights in NNs, the PI algorithm require an admissible controller at the beginning 
of the iteration. However, it is difficult to design one if the dynamics 
of the system are completely unknown. Furthermore, the weight updating method 
is in the least-squares sense, bringing a DT weight controller into 
the actual CT dynamics systems. \cite{Vamvoudakis2013} used the 
gradient descent method to solve the integral-TD (I-TD) equation and 
update the weights in NNs. The algorithm in \cite{Vamvoudakis2013} is 
called synchronous IRL, and it is a partially model-free algorithm for 
solving nonlinear optimal control problems. Synchronous IRL is based 
on the concept of synchronous PI \cite{Vamvoudakis2010}, which can be 
regraded as an extended implementation of general PI (GPI, \cite{Sutton1998}) 
(the value iteration method \cite{Bian2016,Bian2021} is also a special case 
of GPI). The initial admissible policy is not required in synchronous IRL; 
however, full information of the input gain matrix is still required. 
In \cite{Vamvoudakis2017}, a 
Q-learning method for CT linear systems was proposed. This is a completely 
model-free method that is implemented by estimating the Q function instead 
of the original VF.

For the optimal control problems of input-affine nonlinear systems, the application 
of the IRL algorithms is limited by 
several shortcomings. Focusing on these limitations, we propose 
a novel algorithm called synchronous integral Q-learning as a solution. 
Because of the combination of the exploration term and the synchronous 
learning structure, 
the actor and critic NNs can simultaneously and continuously update their 
weights to approximately solve the exploration-HJBE and guarantee the closed-loop 
stability and the convergence of NNs under the PE condition. 
The main contributions of this study are summarised as follows: 

\begin{itemize}
\item
The proposed algorithm is a completely model-free method that can estimate 
the parameters 
without requiring any\emph{ a priori} knowledge (except for the information 
that the system dynamics should be input-affine) or using an identifier NN 
\cite{Bhasin2013}.
\end{itemize}

\begin{itemize}
\item
The initial admissible control policy in traditional PI methods is not 
needed owing to the characteristics of the synchronous IRL algorithm.
\end{itemize}

\begin{itemize}
\item
The hybrid system structure is avoided in this algorithm because the 
weights are updated continuously.
\end{itemize}

The remainder of this paper is organized as follows. In Section \ref{sec2}, the infinite 
horizon optimal control problem in CT input-affine nonlinear systems is formulated. 
The performance index used to evaluate the quality of a controller is presented, 
and the basic offline PI method and the model-free PI algorithm based on IRL and 
exploration are also introduced in this section. Section \ref{sec3} provides the VF 
approximation design of our method and the online weight tuning law based on the 
actor-critic NNs. Then, the closed-loop stability and the convergence of NNs are 
proved. Numerical simulations that show the effectiveness of the proposed method 
are described in Section \ref{sec4}. Finally, Section \ref{sec5} presents the conclusions of the 
study.

For the notations, we use $\Arrowvert X \Arrowvert$ to denote the Euclidean norm, 
$\sqrt{X^{\top}X}$, of the vector or the Frobenius norm, $\sqrt{tr(X^{\top}X)}$, of 
matrix $X$. $X \otimes Y$ denotes the Kronecker product of matrices 
$X$ and $Y$. The function of time, $x(t)$, is also written as $x_t$ or $x$, 
and the function of other variables, $f(x)$, can be written as $f$ in short.

\section{Optimal control problem and PI algorithms}
\label{sec2}

\subsection{Problem formulation}

Let us consider a CT input-affine nonlinear system:
\begin{equation}
\label{eq:system}
\dot{x}=f(x(t))+g(x(t))u(t){\ }{\ }{\ }x(0)=\xi,
\end{equation}
where $x{\ }{\in}{\ }{\mathbb{R}}^n$ and $u{\ }{\in}{\ }{\mathbb{R}}^m$ are the fully 
observable state and the control input, respectively. $\xi$ is the initial state 
of the system. Let us assume that $f(x)+g(x)u$ is Lipschitz on compact set $\Omega$ 
and satisfies $f(0)=0$.

We define the integral form of the infinite horizon performance index as
\begin{equation}
\label{eq:performance}
J(x,u)=\int_0^{\infty}r(x,u)d\tau,
\end{equation}
where $r(x,u)=S(x)+u^{\top}Ru$ with $S(x)>0$ and $R>0$. In this study, our 
goal is to design an optimal control law $u^*$ that stabilizes the system at 
$x=0$ and minimizes the index (\ref{eq:performance}). We use the following 
VF to represent the quality of a policy:
\begin{equation}
\label{eq:VF}
V^\mu(x)=J(x,u)|_{u=\mu(x)},V^\mu(0)=0,
\end{equation}
where $\mu(x)$ is a feedback control law with $\mu(0)=0$, and in the remainder of 
the paper, it is also called policy. With the admissibility of the policy, 
the VF of the policy is well-defined.

\begin{defn}
\label{def:admissible}
(\cite{AbuKhalaf2005}, Admissible control) Policy $\mu(x)$ is said to be admissible 
on $\Omega$, denoted by $\mu{\ }{\in}{\ }\mathcal{A}(\Omega)$, iff the following are 
satisfied:

1) This policy stabilizes (\ref{eq:system}) on $\Omega$, i.e. 
\begin{equation}
\label{eq:stabilize}
\lim_{t{\to}{\infty}}(f+g\mu)d\tau=0.
\end{equation}

2) $V^\mu(\xi)$ is bounded for any state $\xi{\ }{\in}{\ }\Omega$.

Here, $\Omega$ and $\mathcal{A}(\Omega)$ are the admissible region 
of (\ref{eq:system}) and the admissible control set, respectively.
\end{defn}

Let us assume that the admissible control set $\mathcal{A}(\Omega)$ of system 
(\ref{eq:system}) is not empty and $V^\mu{\ }{\in}{\ }\mathcal{C}^1(\Omega)$. According to  
Definition \ref{def:admissible}, it is easy to conclude that there exists an 
optimal control law, $\mu^*(x)$, such that
\begin{equation}
\begin{aligned}
V^{\mu^*}(\xi)&=\min_{\mu(x){\in}\mathcal{A}(\Omega)}\int_0^{\infty}r(x(\tau),u(\tau))d\tau\\
&{\ }{\le}{\ }V^{\mu}(\xi),{\forall}\xi{\ }{\in}{\ }\Omega
\nonumber.
\end{aligned}
\end{equation}

It can be seen clearly that the optimal VF satisfies $V^*(\xi)=V^{\mu^*}(\xi)$. 
In Section \ref{subsec:PI}, we introduce several methods for solving the optimal 
VF. Without special instructions, the problem discussed in this paper is limited 
to a compact set, $x{\ }{\in}{\ }\Omega$.

\subsection{HJBE and PI}\label{subsec:PI}

According to the definition of the VF, the infinitesimal version of (\ref{eq:VF}) 
can be obtained as
\begin{equation}
\label{eq:lyapunov}
0=r(x,\mu(x))+({\nabla}V^\mu)^{\top}(f(x)+g(x)\mu(x)),
\end{equation}
where ${\nabla}V^\mu$ denotes the gradient of $V^\mu$ and (\ref{eq:lyapunov}) is 
called the Lyapunov equation of the system (\ref{eq:system}). From (\ref{eq:VF}) 
and (\ref{eq:lyapunov}), we can infer that
\begin{equation}
\label{eq:V}
V^\mu(x){\ }{\ge}{\ }0,
\end{equation}
\begin{equation}
\label{eq:dotV}
\dot{V}^\mu(x)=-r(x,\mu){\ }{\le}{\ }0.
\end{equation}

Here, $V^\mu(x)$ is regarded as the Lyapunov function of system (\ref{eq:system}). 
The optimal control problem can be converted to an optimization problem under 
the constraint of the state equation. We define the Hamiltonian as follows:
\begin{equation}
\label{eq:hamiltonian}
H(x,u,{\nabla}V^\mu)=r(x,u)+\left({\nabla}V^\mu)^{\top}(f(x)+g(x)u\right),
\end{equation}
where ${\nabla}V^\mu$ is also the Lagrange multiplier for this problem. For the 
optimal policy and its VF, the following Lyapunov equation is satisfied:
\begin{equation}
\label{eq:optimallyap}
H(x,\mu^*,{\nabla}V^{\mu^*})=0.
\end{equation}

The optimal policy can be obtained by minimizing the Hamiltonian:
\begin{equation}
\label{eq:optimalpolicy}
\mu^*=\arg\min_{\mu{\in}\mathcal{A}(\Omega)}H(x,\mu,{\nabla}V^{\mu^*}).
\end{equation}
Owing to the input-affine characteristic of system (\ref{eq:system}), the 
optimal policy can be explicitly given as
\begin{equation}
\label{eq:optimalaffinepolicy}
\mu^*=-\frac{1}{2}R^{-1}g^{\top}(x){\nabla}V^{\mu^*}(x).
\end{equation}

Substituting (\ref{eq:optimallyap}) into (\ref{eq:hamiltonian}), we can obtain the 
well-known HJBE:
\begin{equation}
\begin{aligned}
\label{eq:HJBE}
0&=S(x)+({\nabla}V^{\mu^*})^{\top}(x)f(x)\\
&-\frac{1}{4}({\nabla}V^{\mu^*})^{\top}(x)g(x)R^{-1}g^{\top}(x){\nabla}V^{\mu^*}(x)\\
V&^*(0)=0.
\end{aligned}
\end{equation}
With the linear system dynamics and the quadratic form of the performance index, 
i.e. the linear quadratic regulator (LQR) problem, the HJBE becomes the Riccati 
equation, which is relatively easy to solve. However, in the general nonlinear case, 
it is usually extremely difficult or even not possible to find the solution for the 
HJBE.

PI is a DP algorithm used to iteratively solve the optimal control problem by 
alternately taking two steps, namely, policy evaluation and policy improvement. 
The procedure for offline PI is shown in Algorithm \ref{alg:PI}.

\begin{algorithm}[t]
\label{alg:PI}
\caption{Offline PI}

1. \textbf{Initialization}

Given the initial admissible policy, $\mu_0(x)$, set $i{\ }{\gets}{\ }0$.

2. \textbf{Policy Evaluation}

Solve the Lyapunov equation according to $\mu_i(x)$:
\begin{equation}
\begin{aligned}
\label{eq:policyev}
&H(x,\mu_i(x),{\nabla}V^{\mu_i})=0\\
&V^*(0)=0.
\end{aligned}
\end{equation}

3. \textbf{Policy Improvement}

Update the control policy
\begin{equation}
\label{eq:policyim}
\mu_{i+1}=\arg\min_{\mu{\in}\mathcal{A}(\Omega)}H(x,\mu,{\nabla}V^{\mu_i}).
\end{equation}
For input-affine system (\ref{eq:system}), this policy can be explicitly 
represented as
\begin{equation}
\label{eq:policyimaffine}
\mu_{i+1}=-\frac{1}{2}R^{-1}g^{\top}(x){\nabla}V^{\mu_i}(x).
\end{equation}

4. Set $i{\ }{\gets}{\ }i+1.$

5. Repeat step 2--4 until convergence.
\end{algorithm}

\begin{rem}
In the optimal control problem, the convergence of the PI can be guaranteed if the 
algorithm starts with an initial admissible policy. Under this condition, 
the convergence to the optimal policy and VF has been proven. See 
\cite{AbuKhalaf2005} for the detailed proof.
\end{rem}

With regard to the LQR problem of linear time-invariant systems, Algorithm 
\ref{alg:PI} becomes the Kleinman algorithm \cite{Kleinman1968}. In the case of 
high--order and complex nonlinear systems, the PI algorithm is still difficult 
to implement. The solution to (\ref{eq:policyev}) is often approximated by 
NNs \cite{AbuKhalaf2005}, Galerkin approximation \cite{Beard1997}, and other 
approximation methods. The system dynamics need to be fully known in this 
algorithm.

\subsection{IRL with explorations}

\cite{Vrabie2009} proposed an algorithm framework called the IRL. By integrating 
(\ref{eq:dotV}) into time interval $[t-T,t]$, we can obtain the I-TD equation as
\begin{equation}
\label{eq:ITD}
V^{\mu_i}(x(t-T))=\int_{t-T}^tr(x,\mu_i)d\tau+V^{\mu_i}(x(t)).
\end{equation}

Note that there is no\emph{ a priori} knowledge of the system in 
(\ref{eq:ITD}); the first term on the right-hand side of this equation can be 
collected online. For sufficient groups of integral data, the critic NN and 
the least-squares method can be used to approximate the computation of the solution  
to (\ref{eq:policyev}) and finish the policy evaluation. The policy can be 
updated by using (\ref{eq:policyimaffine}), and thus, the requirement of the known 
system drift dynamics $f(x)$ is dismissed.

Unlike in offline PI, the PE condition is required to guarantee the uniqueness of 
${\nabla}V^{\mu_i}$. However, it cannot re-excite the system when the state 
has been stabilized at the origin. Thus, the convergence to the optimal solution 
may not be guaranteed in real-world implementations. \cite{Lee2012} 
improved the policy evaluation step and solved the input-affine optimal 
control problem. By adding a bounded piecewise continuous nonzero probing 
signal $e_\tau$, we can transform (\ref{eq:system}) into
\begin{equation}
\label{eq:systemexp}
\dot{x}=f(x)+g(x)(u+e).
\end{equation}
The online Lyapunov equation (\ref{eq:ITD}) can be obtained as follows after 
adding the term with $e_\tau$:
\begin{equation}
\label{eq:ITDe1}
\begin{aligned}
V^{\mu_i}(x(t-T))&+\int_{t-T}^t({\nabla}V^{\mu_i})^{\top}g(x)e_{\tau}d\tau\\
&=\int_{t-T}^tr(x,\mu_i)d\tau+V^{\mu_i}(x(t)).
\end{aligned}
\end{equation}

\begin{rem}
Compared with the method in \cite{Vrabie2009}, this method does not require additional 
information on the system dynamics. The designed signal $e_\tau$ is added 
to ensure that the PE condition is satisfied without generating an estimation bias. 
The concept of the probing signal is equivalent to the exploration \cite{Sutton1998} 
in the RL literature.
\end{rem}

By further substituting (\ref{eq:policyimaffine}) into (\ref{eq:ITDe1}), we can 
obtain the following equation:
\begin{equation}
\label{eq:ITDe2}
\begin{aligned}
V^{\mu_i}(x(t-T))&-\int_{t-T}^t2\mu_{i+1}^{\top}Re_{\tau}d\tau\\
&=\int_{t-T}^tr(x,\mu_i)d\tau+V^{\mu_i}(x(t)).
\end{aligned}
\end{equation}

Note that (\ref{eq:ITDe2}) can simultaneously evaluate and improve the present 
policy. During the iteration, no\emph{ a priori} knowledge of the system 
is required. If $e_{\tau}{\ }{\equiv}{\ }0$, (\ref{eq:ITDe2}) is equivalent to 
(\ref{eq:policyev}). The exploration signal can both guarantee the PE condition 
and relax the requirement of $g(x)$, making it a completely model-free 
algorithm. However, two issues exist in this algorithm:

\begin{itemize}
\item
Because of the nature of PI algorithms, both (\ref{eq:ITDe1}) and 
(\ref{eq:ITDe2}) still require an initial admissible policy; this might be 
difficult to implement when the system dynamics are partially or even completely 
unknown.
\end{itemize}

\begin{itemize}
\item
The algorithm updates the VF and the policy based on the batch or recursive 
least-squares method, which brings a DT weight tuning controller to the CT 
system. The hybrid system structure increases the burden on the computing unit.
\end{itemize}

In Section \ref{sec3}, we present a novel algorithm that combines the concepts of 
IRL, exploration, and synchronous RL to solve the aforementioned issues. We call 
this algorithm synchronous integral Q-learning because it is a GPI implementation 
of the algorithm in \cite{Lee2015}.

\section{Synchronous integral reinforcement learning based on explorations}
\label{sec3}

\subsection{Synchronous integral Q-learning}

Eq. (\ref{eq:ITDe2}) shows that the optimal policy and its corresponding VF satisfy 
\begin{equation}
\label{eq:EHJBE}
\begin{aligned}
V^{\mu^*}(x(t-T))&-\int_{t-T}^t2\mu^{*{\top}}Re_{\tau}d\tau\\
&=\int_{t-T}^tr(x,\mu^*)d\tau+V^{\mu^*}(x(t)).
\end{aligned}
\end{equation}

The exploration-HJBE can be approximately solved using the actor-critic NNs. First, 
we consider the VF approximation. We assume that the optimal VF can be denoted as an NN:
\begin{equation}
\label{eq:optVNN}
V^{\mu^*}(x)=w_c^{*{\top}}\phi_c(x)+\varepsilon_c(x),
\end{equation}
where $\phi_c:\mathbb{R}^n{\ }{\to}{\ }\mathbb{R}^{N_c}$, $w_c^*$ and $\varepsilon_c$ are 
the activation function, weight and reconstruction error of the NN, respectively. 
$N_c$ is the number of hidden layers in the critic NN. Because $\varepsilon_c$ is bounded 
on a compact set, the activation function can be selected properly to create a complete 
set of basis functions such that $V^*(x)$ and its gradient
\begin{equation}
\label{eq:optVNNgra}
{\nabla}V^{\mu^*}={\nabla}\phi_c^{\top}w_c^*+{\nabla}\varepsilon_c
\end{equation}
are uniformly approximated \cite{Hornik1990}. According to the Weierstrass high order 
approximation theorem \cite{AbuKhalaf2005}, such a set of basis functions exists
if the VF is sufficiently smooth. Moreover, $\varepsilon_c$ and its gradient 
${\nabla}\varepsilon_c$ are bounded when $N_c$ is a constant and 
$\varepsilon_c{\ }{\to}{\ }0$ uniformly when $N_c{\ }{\to}{\ }\infty$.

Similarly, the optimal policy can be approximated by an actor NN:
\begin{equation}
\begin{aligned}
\label{eq:optuNN}
\mu^*(x)&=-\frac{1}{2}R^{-1}g^\top(x){\nabla}\phi_c^\top(x)w_c^*-\frac{1}{2}R^{-1}g^\top(x){\nabla}\varepsilon_c\\
&=w_a^{*\top}\phi_a(x)+\varepsilon_a(x),
\end{aligned}
\end{equation}
where $\phi_a:\mathbb{R}^n{\ }{\to}{\ }\mathbb{R}^{N_a}$, $w_a^*$ and 
$\varepsilon_a$ are similar to the parameters in the critic NN, which can also enable the  
actor NN uniformly approximate the optimal policy. Using the actor critic NNs, we can define 
the approximation error of (\ref{eq:EHJBE}) as
\begin{equation}
\begin{aligned}
\label{eq:Berror}
\int_{t-T}^t&(S(x)+u^{*\top}Ru^*)d\tau\\
&+w_c^{*\top}\phi_c(x(t))-w_c^{*\top}\phi_c(x(t-T))\\
&+{\rm col}\{w_a^*\}^\top\int_{t-T}^t2\phi_a(x)\otimes(Re_\tau)d\tau{\ }{\equiv}{\ }\varepsilon_B.
\end{aligned}
\end{equation}

By defining the integral reinforcement
\begin{equation}
\rho(x,u)=\int_{t-T}^tr(x_\tau,u_\tau)d\tau,
\end{equation}
we can write (\ref{eq:Berror}) as
\begin{equation}
\varepsilon_B-\rho=W^{*\top}\delta,
\end{equation}
where $W^*=[w_c^{*\top},{\rm col}\{w_a^*\}^\top]^\top$ and
\begin{equation}
\begin{aligned}
\delta&=[\delta_c^\top,\delta_a^\top]^\top\\
&={\rm col}\left\{\phi_c(x)|_{t-T}^t,\int_{t-T}^t2\phi_a(x)\otimes(Re_\tau)d\tau\right\}
\nonumber.
\end{aligned}
\end{equation}

Under the assumption that $f(x)+g(x)u$ is Lipschitz, the residual error 
$\varepsilon_B$ is bounded on a compact set.

\begin{rem}
When $N_c,N_a{\ }{\to}{\ }\infty$, $\varepsilon_B{\ }{\to}{\ }0$ uniformly.
\end{rem}

\subsection{Actor-critic networks and the weight tuning law}
\begin{figure}
\begin{center}
\includegraphics[width=3.47in]{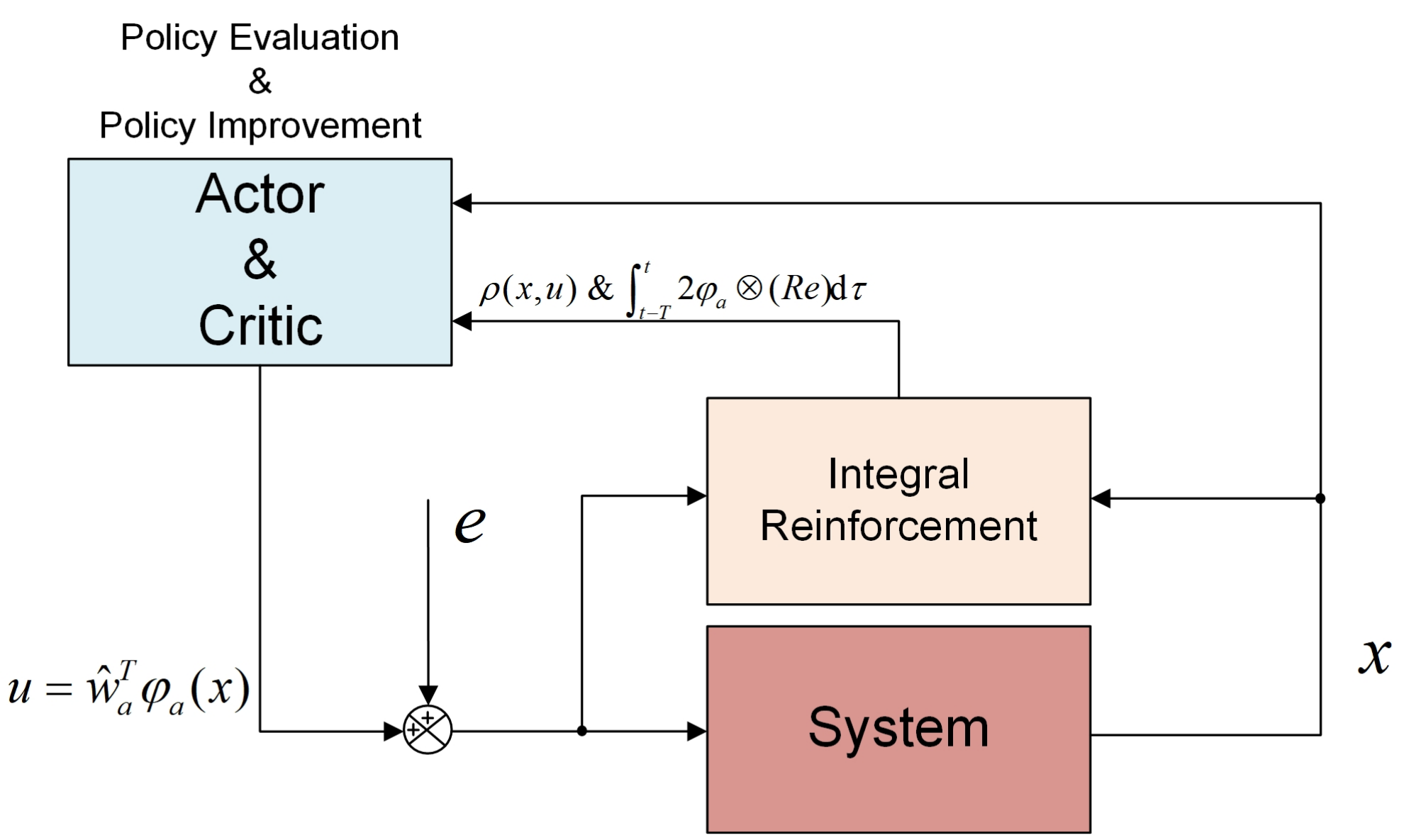}    % The printed column  
\caption{Control scheme of synchronous integral Q-learning algorithm.}  % width is 8.4 cm.
\label{fig1}                                 % Size the figures 
\end{center}                                 % accordingly.
\end{figure}

We use the critic and actor NNs to approximate the optimal VF 
and policy, respectively, according to (\ref{eq:Berror}), and we define the approximate 
exploration-HJBE as
\begin{equation}
\begin{aligned}
\label{eq:appBerror}
\int_{t-T}^t \big( -S(x)-\phi_a^\top(x)w_a^*Rw_a^{*\top}\phi_a(x)+&\varepsilon_{HJB}(x) \big) d\tau\\
&=W^{*\top}\delta,
\end{aligned}
\end{equation}
where $\varepsilon_{HJB}(x)$ is the approximation error arising from the NNs. Because the 
optimal weights $w_c^*$ and $w_a^*$ are unknown, a parameter estimation method 
is required. The estimation of the VF can be obtained as
\begin{equation}
\hat{V}(x)=\hat{w}_c^\top\phi_c(x),
\end{equation}
and the estimation of the policy is
\begin{equation}
\hat{\mu}(x)=\hat{w}_a^\top\phi_a(x),
\end{equation}
where $\hat{w}_c$ and $\hat{w}_a$ are the estimations of the parameters. The 
approximation Bellman error can be obtained from (\ref{eq:Berror}) as
\begin{equation}
E=\hat{W}^\top\delta+\rho,
\end{equation}
where $\hat{W}=[\hat{w}_c^\top,{\rm col}\{\hat{w}_a\}^\top]^\top$. To minimize the squared 
residual error
\begin{equation}
K=\frac{1}{2}E^{\top}E,
\end{equation}
we can use the gradient-based methods to update the weights of both the two NNs. 
By using the normalized gradient descent algorithm \cite{Ioannou2006} and 
(\ref{eq:appBerror}), we can obtain the weights tuning law as
\begin{equation}
\label{eq:tuninglaw}
\dot{\hat{W}}=-\alpha\frac{\partial{K}}{\partial{\hat{W}}}=-\alpha\frac{\delta}{(1+\delta^\top\delta)^2}E,
\end{equation}
where $\alpha>0$ is the learning rate that determines the convergence speed of the 
parameters. The entire control scheme of the algorithm is shown in Fig. \ref{fig1}.

We define $\overline{\delta}=\delta/(1+\delta^\top\delta)$. Before analyzing the 
convergence of the parameters, we need to review the PE conditions in this section.

\begin{defn}
(\cite{Ioannou2006}, PE) At any given time, signal $\overline{\delta}$ is said to be 
persistently excited over interval $[t-T,t]$ if there exist constants $\beta_1>0$ and 
$\beta_2>0$, such that
\begin{equation}
\label{eq:PE}
{\beta}_1I{\ }{\le}{\ }\int_{t-T}^t\overline{\delta}(\tau)\overline{\delta}^\top(\tau)d\tau{\ }{\le}{\ }{\beta}_2I,
\end{equation}
\end{defn}

The PE condition is widely used in adaptive control and system identification methods 
to guarantee the convergence of the parameters.

Defining the estimation error of the weights as $\tilde{W}:=W^*-\hat{W}$, we can express 
the error dynamics as
\begin{equation}
\label{eq:errordynamics}
\left\{
\begin{aligned}
&\dot{\tilde{W}} = -\alpha \overline{\delta} \cdot \overline{\delta}^\top \tilde{W} + \alpha \overline{\delta} \frac{\varepsilon_B}{m_s}\\
&y=\overline{\delta}^\top\tilde{W}
\end{aligned}
\right.,
\end{equation}
where $m_s=1+\delta^\top\delta$. According to (\ref{eq:errordynamics}) and (\ref{eq:PE}), 
we can obtain the following lemma.

\begin{lem}
\label{lemma1}
Assume that the control policy is admissible and that $\overline{\delta}$ is persistently excited 
for all $t>0$. If 
the residual error satisfies $\Arrowvert \varepsilon_B \Arrowvert{\ }{\le}{\ }\varepsilon_{\max}$,
the norm of the estimation error $\Arrowvert \tilde{W} \Arrowvert$ converges exponentially to a 
residual set:
\begin{equation}
\label{eq:lemma}
\tilde{W}{\ }{\le}{\ }\frac{\sqrt{\beta_2 T}}{\beta_1} \{[1 + \eta \beta_2 \alpha] \varepsilon_{\max}\},
\end{equation}
where $\eta$ is a positive constant of the order of 1.
\end{lem}
\textbf{Proof. } See \cite{Vamvoudakis2010}.

Lemma \ref{lemma1} proves that, under the admissible control condition, the weights can 
converge exponentially to a neighborhood of the optimal weights when the reconstruction 
error exists. This is important for evaluating the performance of the algorithm.

We assume the following.

\begin{assum}
\label{assum:bound}
For a given compact set $\Omega{\ }{\in}{\ }\mathbb{R}^n$:

a. $f(\cdot)$ is Lipschitz and $g(\cdot)$ is bounded by a constant
\begin{equation}
\label{eq:assum1a}
{\Arrowvert}f(x){\Arrowvert}<b_f{\Arrowvert}x{\Arrowvert},{\Arrowvert}g(x){\Arrowvert}{\ }{\le}{\ }b_g
\nonumber.
\end{equation}

b. The reconstruction error of the NNs and the gradient of the critic NN error are bounded so that
\begin{equation}
\label{eq:assum1b}
\begin{aligned}
&{\Arrowvert}\varepsilon_c{\Arrowvert}<b_{\varepsilon_c},{\Arrowvert}\varepsilon_c{\Arrowvert}<b_{\varepsilon_c},\\
&{\Arrowvert}\nabla\varepsilon_c{\Arrowvert}<b_{\varepsilon_{cx}}
\nonumber.
\end{aligned}
\end{equation}

c. The activation functions of the NNs and the gradients of the critic NN activation functions are bounded so that
\begin{equation}
\label{eq:assum1c}
\begin{aligned}
&{\Arrowvert}\phi_c(x){\Arrowvert}<b_{\phi_c},{\Arrowvert}\phi_a(x){\Arrowvert}<b_{\phi_a},\\
&{\Arrowvert}\nabla\phi_c(x){\Arrowvert}<b_{\phi_{cx}}
\nonumber.
\end{aligned}
\end{equation}

d. The optimal weights of the NNs are bounded so that
\begin{equation}
\label{eq:assum1d}
{\Arrowvert}W^*{\Arrowvert}<W_{\max}^*
\nonumber.
\end{equation}
\end{assum}

\begin{thm}
\label{thm:UUB}
Let all the assumptions in this paper hold, and let the tuning law and the parameters be 
selected as detailed in the proof. Then, there exists a number $N_0$ such that, for 
the number of hidden layer units of both the two NNs $N_c,N_a>N_0$, the closed loop 
system state and the NN approximation error $\tilde{W}$ are uniformly ultimately 
bounded (UUB).
\end{thm}
\textbf{Proof. }See Appendix \ref{app:A}.

\subsection{Implementation of the algorithm for LQR problems}
Let us consider the widely studied CT LQR problem, i.e. $f(x)=Ax,g(x)=B$, where $A$ and 
$B$ are matrices that do not depend on $x$. Specially, we define the performance index 
as $J=x^{\top}Sx+u^{\top}Ru$ with $S>0$. According to the basic LQR theory, the optimal 
VF is quadratic to $x$ and the optimal policy is the linear feedback control of $x$
\begin{equation}
V^*=w_c^{*\top}\phi_c(x)=x^{\top}P^*x
\nonumber,
\end{equation}
\begin{equation}
\mu^*=w_a^{*\top}x=-R^{-1}B^{\top}P^*x
\nonumber,
\end{equation}
where $\phi_c(x)=x \otimes x$. The exploration-HJBE (\ref{eq:EHJBE}) in the LQR problem 
becomes
\begin{equation}
\label{eq:linearEHJBE}
\int_{t-T}^t(-x^{\top}Sx-x^{\top}w_a^*Rw_a^{*\top}x)d\tau=W^{*\top}\delta.
\end{equation}
Note that the approximation error $\varepsilon_{HJB}(x)$ does not occur in 
(\ref{eq:linearEHJBE}). Similarly, the approximation NNs can be written as
\begin{equation}
\hat{V}(x)=\hat{w}_c^{\top}(x \otimes x),\hat{\mu}(x)=\hat{w}_a^{\top}x
\nonumber.
\end{equation}
Then, using the weight tuning law (\ref{eq:tuninglaw}), we can solve the LQR problem  
online. The policy is also globally optimal for linear systems, and the approximation 
error is guaranteed to converge exponentially to zero owing to the non-existence of 
the reconstruction error of the NNs.

\begin{rem}
For Linear systems, the exploration can be chosen as a sum of sinusoidal signals 
that have sufficient richness (the number of the frequency components must be larger 
than or equal to the number of estimated parameters) to satisfy the PE condition. 
However, in nonlinear problems, no verifiable method exists to ensure that 
\cite{Vamvoudakis2013}.
\end{rem}

\begin{rem}
After the exploration signal is added, both the actor and the critic NN can update their 
weights by solving the same equation and the state-value function is approximated 
in this algorithm instead of directly estimating the Q function. Thus, the proposed 
method is different from the Q-learning approaches in 
\cite{Vamvoudakis2017,Lee2020,Chen2019}.
\end{rem}

\section{Numerical simulations}
\label{sec4}

To show the effectiveness of the proposed method, we set a second order 
nonlinear system as a benchmark, which has been used in several studies 
\cite{Vamvoudakis2010,Bhasin2013,Lee2015}. The system dynamics are as follows:
\begin{equation}
f(x)=\left[
\begin{array}{c}
-x_1+x_2\\
-0.5x_1-0.5x_2(1-(\cos(2x_1)+2)^2)
\end{array}
\right],
\end{equation}
\begin{equation}
g(x)=\left[
\begin{array}{c}
0\\
\cos(2x_1)+2
\end{array}
\right].
\end{equation}
The cost function is selected as
\begin{equation}
S(x)=x_1^2+x_2^2,R=1
\nonumber.
\end{equation}

According to the converse HJB approach \cite{Nevistic1996}, the optimal VF and 
policy can be respectively obtained as
\begin{equation}
\label{eq:optV}
V^*(x)=\frac{1}{2}x_1^2+x_2^2
\end{equation}
and
\begin{equation}
\label{eq:optMu}
\mu^*(x)=-(\cos(2x_1)+2)x_2.
\end{equation}
Here, we present two cases of this example to show the approximation performance 
of the two NNs.

\subsection{Case 1: exact parameterization}
Now, let us assume that the VF and the policy are parameterized exactly. In this case, 
the algorithm is used to estimate the parameters in a grey-box fashion. We choose 
the activation function as follows:
\begin{equation}
\begin{aligned}
&\phi_c(x)=[x_1^2,x_1x_2,x_2^2]^\top,\\
&\phi_a(x)=[x_1\cos(2x_1),x_1,x_2\cos(2x_1),x_2]^\top
\nonumber.
\end{aligned}
\end{equation}
The optimal weights can be obtained from (\ref{eq:optV}) and(\ref{eq:optMu}) and 
are
\begin{equation}
\begin{aligned}
\label{eq:optW}
&w_c^*=[0.5,0,1]^\top,\\
&w_a^*=[0,0,-1,-2]^\top.
\end{aligned}
\end{equation}

We choose the initial state as $x(0)=[0,0]^\top$ and the initial weights of the NNs as 
$\hat{w}_c(0)=[1,1,1]^\top$ and $\hat{w}_a(0)=[0.5,-0.5-0.5,-0.5]^\top$. The learning
rate is set as $\alpha=1000$.

\begin{figure}
\begin{center}
\includegraphics[width=3.47in]{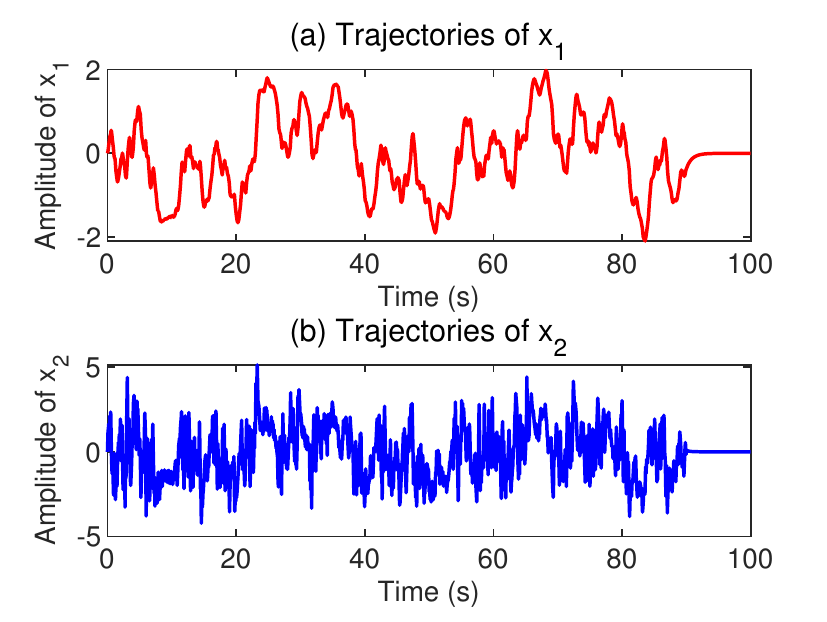}    % The printed column  
\caption{Case 1: trajectories of (a) $x_1$ and (b) $x_2$.}  % width is 8.4 cm.
\label{fig2}                                 % Size the figures 
\end{center}                                 % accordingly.
\end{figure}

The design of the exploration signal determines the level of excitation, which also
affects the performance of the algorithm. In this case we choose the exploration 
signal as
\begin{equation}
e(t)=\sum_{k=1}^{100}\sin(\omega_kt)
\nonumber,
\end{equation}
where $\omega_k$ is uniformly sampled from $[-50,50]$. The exploration is added to 
$t{\ }{\in}{\ }[0,90]$. After 90 s, the exploration is ended and the simulation 
stops at $t_f=100$ s. The length of the sampling interval is $T=0.025$ s. The 
trajectories of $x_1$ and $x_2$ are shown in Fig. \ref{fig2}. After the exploration 
is stopped, the state can be stabilized near the origin.

\begin{figure}
\begin{center}
\includegraphics[width=3.47in]{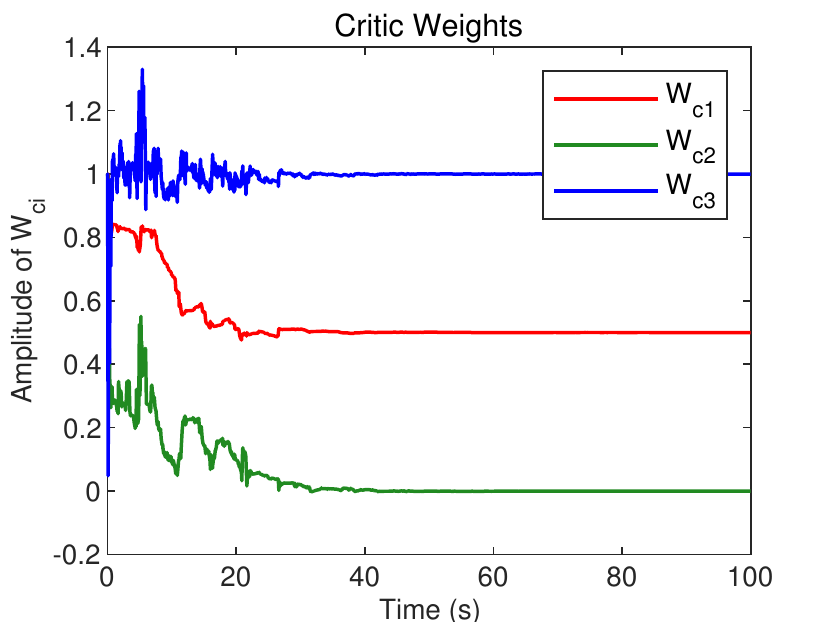}    % The printed column  
\caption{Case 1: evolution of the critic weights.}  % width is 8.4 cm.
\label{fig3}                                 % Size the figures 
\end{center}                                 % accordingly.
\end{figure}

\begin{figure}
\begin{center}
\includegraphics[width=3.47in]{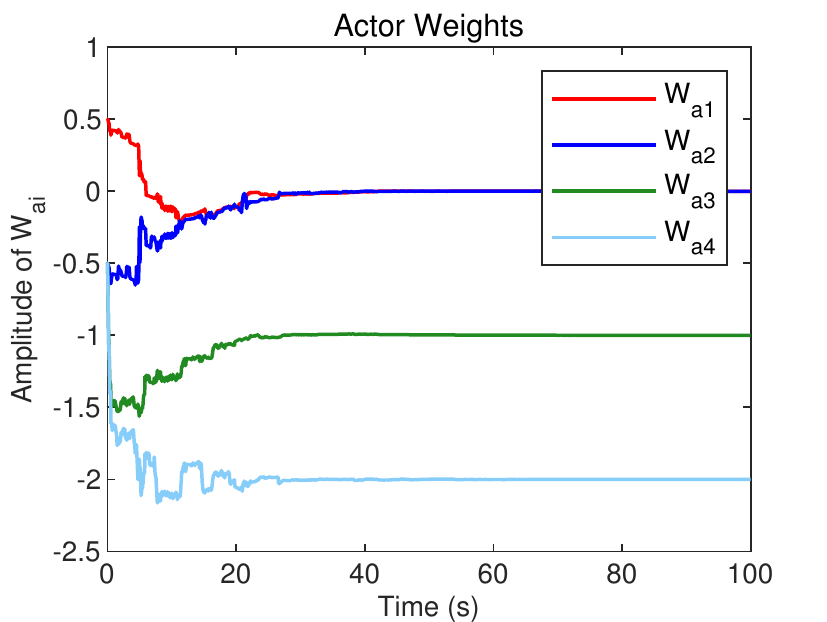}    % The printed column  
\caption{Case 1: evolution of the actor weights.}  % width is 8.4 cm.
\label{fig4}                                 % Size the figures 
\end{center}                                 % accordingly.
\end{figure}

\begin{figure}
\begin{center}
\includegraphics[width=3.47in]{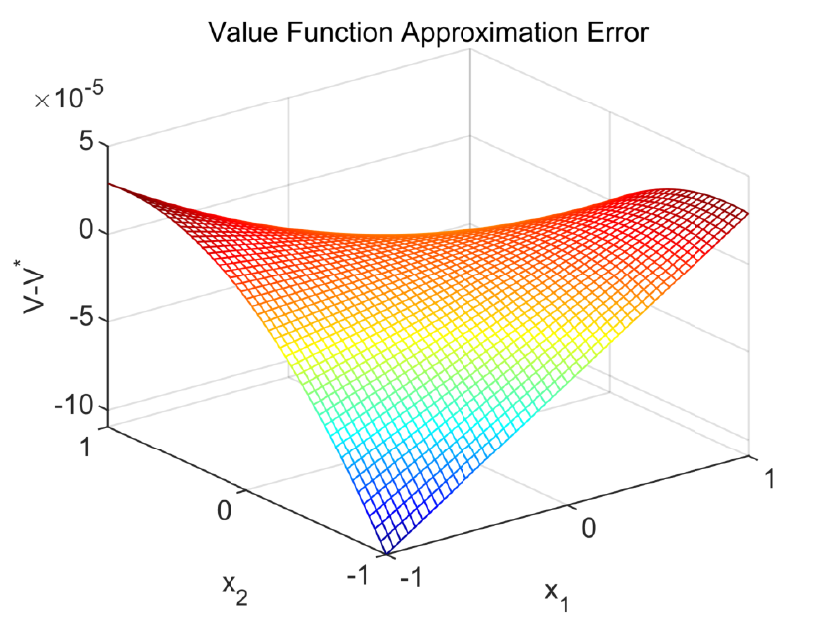}    % The printed column  
\caption{Case 1: approximation error of the critic network.}  % width is 8.4 cm.
\label{fig5}                                 % Size the figures 
\end{center}                                 % accordingly.
\end{figure}

\begin{figure}
\begin{center}
\includegraphics[width=3.47in]{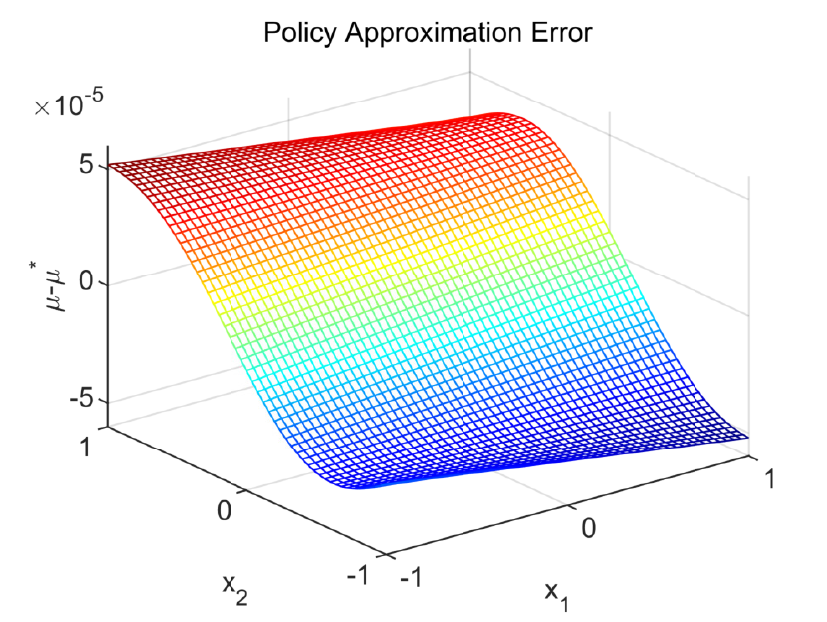}    % The printed column  
\caption{Case 1: approximation error of the actor network.}  % width is 8.4 cm.
\label{fig6}                                 % Size the figures 
\end{center}                                 % accordingly.
\end{figure}

As shown in Figs. \ref{fig3} and \ref{fig4}, all the weights in the critic and 
actor NNs are close to the optimal value. After 100 s of training, the weights 
of the two NNs converge to
\begin{equation}
\begin{aligned}
&\hat{w}_c(t_f)=[0.5000,-0.0001,1.0000]^\top,\\
&\hat{w}_a(t_f)=[0.0000,0.0001,-1.0000,-2.0000]^\top
\nonumber,
\end{aligned}
\end{equation}
which are extremely close to the optimal value (\ref{eq:optW}). Figs. \ref{fig5} 
and \ref{fig6} show the approximation errors of the critic and actor NNs, 
respectively. In the region of $x_1,x_2{\ }{\in}{\ }[-1,1]$, the maximum 
approximation error of the VF is approximately $10^{-4}$ and that of the policy is 
approximately $5{\times}10^{-5}$, indicating the excellent approximation 
performance of the trained NN.

\subsection{Case 2: fully unknown dynamics}

In case 1, the policy is assumed to satisfy the condition of the exact 
parameterization, which cannot be generalized to the case in which the information 
on the system is completely unknown. In case 2, we choose the following 
activation function to approximate the optimal policy:
\begin{equation}
\phi_a(x)=[x_1,x_1^2,...,x_1^5,x_2,x_1x_2,...,x_1^4x_2]^\top
\nonumber.
\end{equation}

In the neighborhood of the origin, the optimal weight of the policy can be 
obtained as
\begin{equation}
\phi_a(x)=[0,0,0,0,0,-3,0,2,0,-2/3]^\top
\nonumber
\end{equation}
because the Taylor expansion of $\cos(2x_1)+2$ at $x_1=0$ is
\begin{equation}
\cos(2x_1)+2=3-2x_1^2+\frac{2}{3}x_1^4+\mathcal{O}(x_1^6)
\nonumber.
\end{equation}

After the training, the weights converge to
\begin{equation}
\begin{aligned}
&\hat{w}_c(t_f)=[0.5007,0.0011,0.9997]^\top,\\
&\hat{w}_a(t_f)=[-0.0021,-0.0007,0.0040,0.0016,-0.0001,\\
&-3.0011,0.0023,1.9986,0.0040,-0.5758]^\top
\nonumber.
\end{aligned}
\end{equation}

\begin{figure}
\begin{center}
\includegraphics[width=3.47in]{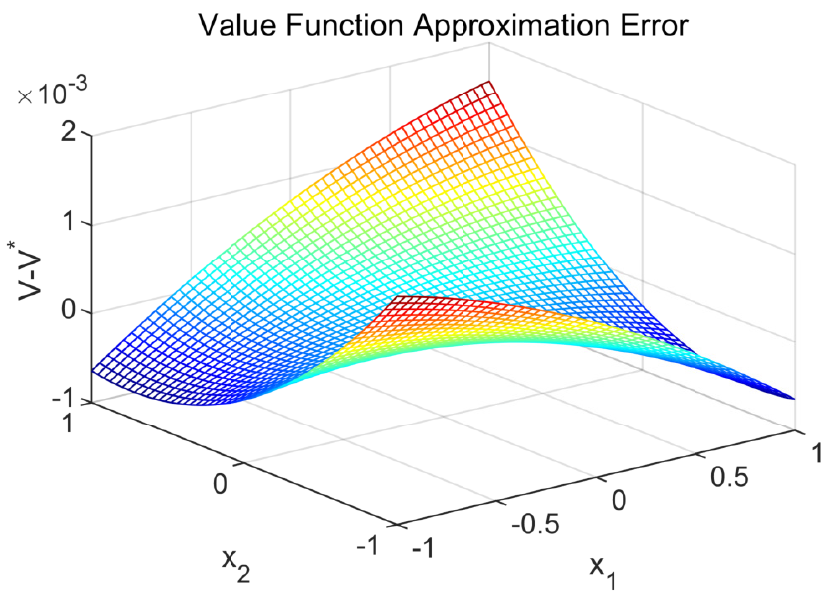}    % The printed column  
\caption{Case 2: approximation error of the critic network.}  % width is 8.4 cm.
\label{fig7}                                 % Size the figures 
\end{center}                                 % accordingly.
\end{figure}

\begin{figure}
\begin{center}
\includegraphics[width=3.47in]{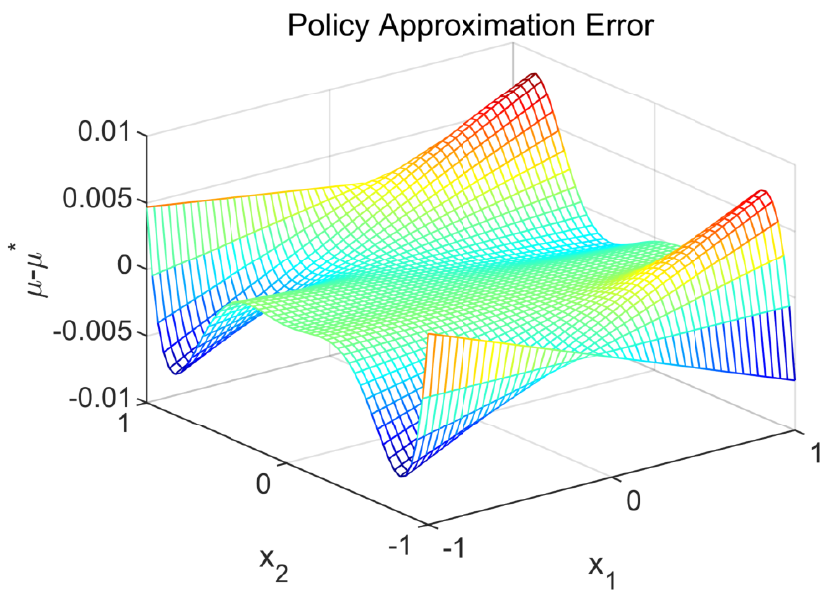}    % The printed column  
\caption{Case 2: approximation error of the actor network.}  % width is 8.4 cm.
\label{fig8}                                 % Size the figures 
\end{center}                                 % accordingly.
\end{figure}

The approximation errors of the optimal VF and policy are shown in Figs. \ref{fig7}
and \ref{fig8}, respectively. The errors of both the NNs are less than $10^{-2}$.

\begin{rem}
Because of the existence of the reconstruction error and the different structures of the 
actor NN between case 1 and case 2, the results of case 2 are worse but can show 
the convergence of the algorithm.
\end{rem}

\begin{rem}
Compared with similar methods, the algorithm proposed in this paper does not require 
an extra identifier NN \cite{Bhasin2013}. In addition, case 1 shows that our method obtains 
a smaller approximation error than does the method in \cite{Vamvoudakis2010} for the 
same training time.
\end{rem}

\section{Conclusions}
\label{sec5}
In this paper, we presented a novel algorithm using the concepts of IRL and synchronous 
RL to solve the CT optimal control problems. It does not require any\emph{ a priori} 
knowledge or an identifier NN. Moreover, an admissible control is not needed for its 
implementation. The design of the exploration to achieve safe 
learning is a meaningful future research direction. In \cite{Lee2015}, the invariant 
exploration method is implemented in the PI algorithm; however, it has not been proven 
to guarantee stability in the GPI method. The extension of our method to multi-agent 
or nonaffine nonlinear control problems is also worth investigating. In addition, it is 
important to explore the application of the proposed method to real-world high-order 
systems, e.g. in designing the controller for robots and aircraft.

\appendix[Proof of Theorem \ref{thm:UUB}]\label{app:A}

We define the approximation errors $\tilde{w}_c=w_c^*-\hat{w}_c$ and $\tilde{w}_a=w_a^*-\hat{w}_a$ 
and consider the Lyapunov function,
\begin{equation}
\label{eq:A1}
L=V^*(x)+\frac{1}{2}\tilde{W}^\top\tilde{W},t{\ }{\ge}{\ }0.
\end{equation}
The derivative of (\ref{eq:A1}) to time $t$ is
\begin{equation}
\label{eq:A2}
\dot{L}=\dot{V}^*(x)+\frac{1}{2}\tilde{W}^\top\dot{\tilde{W}},
\end{equation}
Substituting the error dynamics (\ref{eq:errordynamics}), we can obtain the derivative as
\begin{equation}
\label{eq:A3}
\dot{L}={\nabla}V^{*\top}(x)(f(x)+g(x)\hat{w}_a^\top\phi_a(x)+e)-\alpha\tilde{W}^\top\overline{\delta} \cdot \overline{\delta}^\top\tilde{W}.
\end{equation}

Eq. (\ref{eq:A3}) can be written as two terms, i.e. $\dot{L}=L_1+L_2$, where
\begin{equation}
\label{eq:A4}
L_1={\nabla}V^{*\top}(x)\left(f(x)+g(x)\hat{w}_a^\top\phi_a(x)+e \right),
\end{equation}
\begin{equation}
\label{eq:A5}
L_2=-\alpha\tilde{W}^\top\overline{\delta} \cdot \overline{\delta}^\top\tilde{W}.
\end{equation}
The first term is
\begin{equation}
\begin{aligned}
\label{eq:A6}
L_1&=w_c^{*\top}\left({\nabla}\phi_c(x)f(x)+{\nabla}\phi_c(x)g(x)w_a^{*\top}\phi_a(x))\right.\\
&\left.-{\nabla}\phi_c(x)g(x)\tilde{w}_a^\top\phi_a(x)+{\nabla}\phi_c(x)g(x)e\right)+\varepsilon_1(x),
\end{aligned}
\end{equation}
where
\begin{equation}
\label{eq:A7}
\begin{aligned}
\varepsilon_1(x)&={\nabla}\varepsilon_c^\top\left(f(x)+g(x)w_a^{*\top}\phi_a(x)+g(x)e\right.\\
&\left.-g(x)\tilde{w}_a^\top\phi_a(x)\right).
\end{aligned}
\end{equation}
By substituting the exploration-HJBE (\ref{eq:EHJBE}), we can obtain
\begin{equation}
\begin{aligned}
\label{eq:A8}
L_1&=\left(-S(x)-\phi_a^\top w_a^*Rw_a^{*\top}\phi_a+\varepsilon_{HJB}\right.\\
&\left.-w_c^{*\top}{\nabla}\phi_c(x)g(x)\tilde{w}_a^\top\phi_a(x)\right)+\varepsilon_1(x).
\end{aligned}
\end{equation}

Because $S(x)>0$, there exists matrix $q$ on $\Omega$ such that $x^{\top}qx<S(x)$. 
Substituting $q$ and the relationship between the two NNs, we can write the first term 
of $\dot{L}$ as
\begin{equation}
\begin{aligned}
\label{eq:A9}
L_1&{\ }{\le}{\ }\left(-x^{\top}qx-\phi_a^\top w_a^*Rw_a^{*\top}\phi_a+\varepsilon_{HJB}\right.\\
&\left.+(2\phi_a^\top w_a^*R+2\varepsilon_a^\top R+\nabla \varepsilon_c^\top g(x))\tilde{w}_a^\top \phi_a \right)+\varepsilon_1(x).
\end{aligned}
\end{equation}

Using Young's inequality, we can express (\ref{eq:A9}) as
\begin{equation}
\begin{aligned}
\label{eq:A10}
L_1&{\ }{\le}{\ }-\sigma_{\min}(q)\Arrowvert x \Arrowvert^2+\phi_a^\top \tilde{w}_aR\tilde{w}_a^\top \phi_a+\varepsilon_{HJB}\\
&+(2\varepsilon_a^\top R+{\nabla}\varepsilon_c^\top g(x))\tilde{w}_a^\top \phi_a+\varepsilon_1(x).
\end{aligned}
\end{equation}

We select proper $N_0$ such that $\sup_{x{\in}\Omega}\Arrowvert \varepsilon_{HJB} \Arrowvert<\varepsilon$. 
According to (\ref{eq:A7}) and Assumption \ref{assum:bound}, we can obtain
\begin{equation}
\begin{aligned}
\label{eq:A11}
L_1&{\ }{\le}{\ }-\sigma_{\min}(q)\Arrowvert x \Arrowvert^2+\sigma_{\max}(R)\Arrowvert \tilde{w}_a^\top \phi_a \Arrowvert^2\\
&+2b_{\varepsilon_a}\sigma_{\max}(R) \Arrowvert \tilde{w}_a^\top \phi_a \Arrowvert+\varepsilon\\
&+b_{\varepsilon_c}\left(b_f \Arrowvert x \Arrowvert+b_gb_{\phi_a}\Arrowvert w_a^* \Arrowvert+b_g\Arrowvert e \Arrowvert\right).
\end{aligned}
\end{equation}

By using the characteristics of the norm, we can write (\ref{eq:A11}) as
\begin{equation}
\begin{aligned}
\label{eq:A12}
L_1&{\ }{\le}{\ }-\sigma_{\min}(q)\Arrowvert x \Arrowvert^2+\sigma_{\max}(R)b_{\phi_a}^2\Arrowvert \tilde{W} \Arrowvert^2\\
&+2b_{\varepsilon_a}\sigma_{\max}(R)b_{\phi_a} \Arrowvert \tilde{W} \Arrowvert + b_{\varepsilon_c}b_f\Arrowvert x \Arrowvert\\
&+\varepsilon+b_{\varepsilon_c}\left(b_gb_{\phi_a}\Arrowvert w_a^* \Arrowvert+b_g\Arrowvert e \Arrowvert\right).
\end{aligned}
\end{equation}

According to the proof of Lemma \ref{lemma1}, the second term satisfies
\begin{equation}
\label{eq:A13}
L_2{\ }{\le}{\ }-\alpha\left\Arrowvert \frac{\delta}{m_s} \right\Arrowvert^2 \Arrowvert \tilde{W} \Arrowvert^2+\alpha \left\Arrowvert \frac{\delta}{m_s} \right\Arrowvert \left\Arrowvert \frac{\varepsilon_2}{m_s} \right\Arrowvert \Arrowvert \tilde{W} \Arrowvert,
\end{equation}
where
\begin{equation}
\label{eq:A14}
\varepsilon_2(x)=\nabla \varepsilon_c^\top \left(f(x)+g(x)w_a^{*\top}\phi_a(x)+g(x)e\right). 
\end{equation}

We add (\ref{eq:A12}) and (\ref{eq:A13}) and then substitute (\ref{eq:A14}) and  
inequality $\left\Arrowvert \frac{\delta}{m_s^2} \right\Arrowvert<1$ so that
\begin{equation}
\begin{aligned}
\label{eq:A15}
\dot{L}&{\ }{\le}{\ }-\sigma_{\min}(q)\Arrowvert x \Arrowvert^2\\
&+\left(\sigma_{\max}(R)b_{\phi_a}^2-\alpha\left\Arrowvert \frac{\delta}{m_s} \right\Arrowvert^2\right)\Arrowvert \tilde{W} \Arrowvert^2\\
&+ b_{\varepsilon_{cx}}b_f\Vert x \Vert \Arrowvert \tilde{W} \Arrowvert\\
&+ b_{\varepsilon_c}b_f\Arrowvert x \Arrowvert\\
&+\left(2b_{\varepsilon_a}\sigma_{\max}(R)b_{\phi_a}+ \alpha b_{\varepsilon_{cx}}b_g(b_{\phi_a}\Vert w_a^* \Vert+\Vert e \Vert)\right) \Arrowvert \tilde{W} \Arrowvert \\
&+\varepsilon+b_{\varepsilon_c}\left(b_gb_{\phi_a}\Arrowvert w_a^* \Arrowvert+b_g\Arrowvert e \Arrowvert\right).
\end{aligned}
\end{equation}

Let
\begin{equation}
\begin{aligned}
a=\sigma_{\max} (R)b_{\phi_a}^2-\alpha\left\Arrowvert \frac{\delta}{m_s} \right\Arrowvert^2,\\
c=b_{\varepsilon_c}b_g\left(b_{\phi_a}\Arrowvert w_a^* \Arrowvert+\Arrowvert e \Arrowvert\right)
\nonumber,
\end{aligned}
\end{equation}
and
\begin{equation}
d=\left[
\begin{array}{c}
b_{\varepsilon_c}b_f\\
2b_{\varepsilon_a}\sigma_{\max}(R)b_{\phi_a}+ \alpha b_{\varepsilon_{cx}}b_g(b_{\phi_a}\Vert w_a^* \Vert+\Vert e \Vert)
\end{array}
\right]
\nonumber.
\end{equation}

\begin{equation}
\tilde{Z}=\left[
\begin{array}{c}
\Vert x \Vert\\
\Vert \tilde{W} \Vert
\end{array}
\right]
\nonumber,
\end{equation}
Then, (\ref{eq:A15}) becomes
\begin{equation}
\label{eq:A16}
\dot{L}{\ }{\le}{\ }-\tilde{Z}^\top M \tilde{Z}+d^\top\tilde{Z} +c+\varepsilon,
\end{equation}
where
\begin{equation}
M=\left[
\begin{array}{cc}
\sigma_{\max}(q) & -\frac{b_{\varepsilon_{cx}}b_f}{2}\\
-\frac{b_{\varepsilon_{cx}}b_f}{2} & -a
\end{array}
\right]
\nonumber.
\end{equation}

To let $M$ be a positive definite matrix, we choose a sufficiently large learning rate 
$\alpha$ if $\Vert \delta \Vert \neq 0$. The norm of $\delta$ can easily maintain a 
non-zero value under the PE assumption and a proper value of $T$ during the learning 
phase.
\begin{equation}
\label{eq:A17}
det(M)=-a\sigma_{\max}(q)-\frac{b_{\varepsilon_{cx}}^2b_f^2}{4}>0.
\end{equation}

Then (\ref{eq:A16}) becomes
\begin{equation}
\label{eq:A18}
\dot{L}{\ }{\le}{\ }-\sigma_{\min}(M)\Vert \tilde{Z} \Vert^2+\Vert d \Vert \Vert \tilde{Z} \Vert +c+\varepsilon,
\end{equation}

According to (\ref{eq:A18}), the Lyapunov function is negative if
\begin{equation}
\label{eq:A19}
\Arrowvert \tilde{Z} \Arrowvert > \frac{\Arrowvert d \Arrowvert}{2\sigma_{\min} (M)}+\sqrt{\frac{\Arrowvert d \Arrowvert^2}{4\sigma_{\min}^2(M)}+\frac{c+\varepsilon}{\sigma_{\min}(M)}}{\ }{\equiv}{\ }B_Z.
\end{equation}
The inequality shows that $\dot{L}$ is negative if $L$ exceeds a certain 
bound. Then, according to the Lyapunov analysis, the state and the weights 
are UUB. Under the ideal condition, i.e. $N_c,N_a{\ }{\to}{\ }\infty$ or 
both the optimal VF and the corresponding policy are under the exact  
parameterization assumption, and the state and the approximation error are 
stabilized at the origin.

This completes the proof.

% or


%\appendices

%\section{}

\begin{comment}
\section*{Acknowledgment}

The authors would like to thank...
\end{comment}

\ifCLASSOPTIONcaptionsoff
  \newpage
\fi



%\bibliographystyle{IEEEtran}
\bibliographystyle{IEEEtran}
\bibliography{IEEEabrv,root}



%\newpage


%\vfill



\end{document}